\documentclass[a4paper,zpreprint,zeusbst,british]{zeus_paper}
\usepackage{zeusdet_hepunits}
\usepackage{hangcaption}
\usepackage{subfigure}
\usepackage{amsmath}
\usepackage[left]{lineno}
 
\chardef\usc=95
\chardef\til=126
\catcode`\@=11 
\DeclareRobustCommand\xdotspace{\futurelet\@let@token\@xdotspace}
\def\@xdotspace{%
  \ifx\@let@token.\else
  \ifx\@let@token\bgroup.\else
  \ifx\@let@token\egroup.\else
  \ifx\@let@token\/.\else
  \ifx\@let@token\ .\else
  \ifx\@let@token~.\else
  \ifx\@let@token!.\else
  \ifx\@let@token,.\else
  \ifx\@let@token:.\else
  \ifx\@let@token;.\else
  \ifx\@let@token?.\else
  \ifx\@let@token/.\else
  \ifx\@let@token'.\else
  \ifx\@let@token).\else
  \ifx\@let@token-.\else
  \ifx\@let@token\@xobeysp.\else
  \ifx\@let@token\space.\else
  \ifx\@let@token\@sptoken.\else
   .\space
   \fi\fi\fi\fi\fi\fi\fi\fi\fi\fi\fi\fi\fi\fi\fi\fi\fi\fi}
\catcode`\@=12 

\newcommand{\stru}[2]{%
   \relax\ifmmode\hbox{\vrule height#1 depth#2 width0pt}%
   \else\vrule height#1 depth#2 width0pt\fi}

\newcommand{\Ronum}[1]{\uppercase\expandafter{\romannumeral#1}}
\newcommand{\ronum}[1]{\expandafter{\romannumeral#1}}
\DeclareRobustCommand{\LaTeXZ}{%
  \LaTeX\kern-.05em4\kern-.1em
  {\raisebox{-0.2ex}{$\scriptstyle\text{ZEUS}$}}\xspace}

\newcommand{\slashfrac}[2]{%
  \raisebox{0.5ex}{\ensuremath #1}\kern-0.12em/\kern-0.08em
  \raisebox{-.8ex}{\ensuremath #2}}

\newcommand{\sqr}[3]{%
    {\vcenter{\hrule height.#3ex\hbox{\vrule width.#2ex height#1ex
     \kern#1ex\vrule width.#3ex}\hrule height.#2ex}}}

\catcode`\@=11 
\newcommand{\parenbar}{\mathpalette\p@renb@r}
\def\p@renb@r#1#2{\vbox{%
  \ifx#1\scriptscriptstyle \dimen@.7em\dimen@ii.2em\else
  \ifx#1\scriptstyle \dimen@.8em\dimen@ii.25em\else
  \dimen@1em\dimen@ii.4em\fi\fi \offinterlineskip
  \ialign{\hfill##\hfill\cr
    \vbox{\hrule width\dimen@ii}\cr
    \noalign{\vskip-.3ex}%
    \hbox to\dimen@{$\mathchar300\hfil\mathchar301$}\cr
    \noalign{\vskip-.3ex}%
    $#1#2$\cr}}}
\catcode`\@=12 

\ifzeusunit
  
\else
  
\fi

\newcommand{\IP}{{\rm I$\kern-0.01667em$P}\xspace}

\mathchardef\qsm=63
\mathchardef\pls=43
\mathchardef\mns=512
\mathchardef\plm=518
\mathchardef\eql=61
\mathchardef\smallleft=300
\mathchardef\smallright=301
\mathchardef\les=316
\mathchardef\gre=318
\mathchardef\leq=532
\mathchardef\grq=533

\catcode`\@=11 
\newcounter{pict@width}
\newcounter{pict@height}
\newlength{\pict@scale}
\newcommand{\psfigadd}[4]{%
\setcounter{pict@width}{1*\ratio{#2+\pict@scale/2}{\pict@scale}}
\setcounter{pict@height}{1*\ratio{#3+\pict@scale/2}{\pict@scale}}
\setlength{\unitlength}{\pict@scale}
\hbox to #2{\hspace{-\fill}\begin{picture}(\thepict@width,\thepict@height)
\put(0,0){\psfig{figure=#1,width=#2,height=#3,clip=}}
\SetScale{0.283466457}
\SetWidth{1.763889}
{#4}
\end{picture}}
}
\newcounter{pict@widthfst}
\newcounter{pict@widthscd}
\newcounter{pict@widthtot}
\newcommand{\psfigaddtwo}[7]{%
\setcounter{pict@widthfst}{1*\ratio{#2+\pict@scale/2}{\pict@scale}}
\setcounter{pict@widthscd}{1*\ratio{#2+#4+\pict@scale/2}{\pict@scale}}
\setcounter{pict@widthtot}{1*\ratio{#2+#4+#6+\pict@scale/2}{\pict@scale}}
\setcounter{pict@height}{1*\ratio{#3+\pict@scale/2}{\pict@scale}}
\setlength{\unitlength}{\pict@scale}
\hbox{\hspace{-\fill}\begin{picture}(\thepict@widthtot,\thepict@height)
\put(0,0){\psfig{figure=#1,width=#2,height=#3,clip=}}
\put(\thepict@widthscd,0){\psfig{figure=#5,width=#6,height=#3,clip=}}
\SetScale{0.283466457}
\SetWidth{1.763889}
{#7}
\end{picture}}
}
\newcommand{\psfigror}[4]{%
\setcounter{pict@width}{1*\ratio{#2+\pict@scale/2}{\pict@scale}}
\setcounter{pict@height}{1*\ratio{#3+\pict@scale/2}{\pict@scale}}
\setlength{\unitlength}{\pict@scale}
\hbox{\begin{picture}(\thepict@width,\thepict@height)
\put(0,\thepict@height){\psfig{figure=#1,width=#3,height=#2,clip=,angle=270}}
\SetScale{0.283466457}
\SetWidth{1.763889}
{#4}
\end{picture}}
}
\newcommand{\psfigrol}[4]{%
\setcounter{pict@width}{1*\ratio{#2+\pict@scale/2}{\pict@scale}}
\setcounter{pict@height}{1*\ratio{#3+\pict@scale/2}{\pict@scale}}
\setlength{\unitlength}{\pict@scale}
\hbox{\begin{picture}(\thepict@width,\thepict@height)
\put(0,0){\psfig{figure=#1,width=#3,height=#2,clip=,angle=90}}
\SetScale{0.283466457}
\SetWidth{1.763889}
{#4}
\end{picture}}
}
\catcode`\@=12 
\newlength\listtextwidth

\catcode`\@=11 
\newlength{\@tabfninsert}
\newlength{\@tabfnwidth}
\newcommand{\tabfootnote}[2]{%
  \setlength{\@tabfninsert}{0.8em}
  \setlength{\@tabfnwidth}{\textwidth}
  \addtolength{\@tabfnwidth}{-\@tabfninsert}
  \addtolength{\@tabfnwidth}{-0.4em}
  \noindent\makebox[\@tabfninsert][r]{\footnotesize$^{#1}$\hfil}\hfill%
  \parbox[t]{\@tabfnwidth}{\footnotesize #2\hfill}}
\catcode`\@=12 

\newcommand{\bvec}[1]{\mbox{\boldmath $#1$}}

\newcommand{\figref}[1]{Fig.~\ref{#1}}
\newcommand{\figsref}[1]{Figs.~\ref{#1}}

\newcommand{\eref}[1]{Eq.~(\ref{#1})}

\def\citeCTD{{\cite{%
nim:a279:290,*npps:b32:181,*nim:a338:254%
}}\xspace}
\def\citeMVD{{\cite{%
nim:a581:656%
}}\xspace}
\def\citeSTT{{\cite{%
nim:a535:191%
}}\xspace}
\def\citeCAL{{\cite{%
nim:a309:77,*nim:a309:101,*nim:a321:356,*nim:a336:23%
}}\xspace}
\def\citeHES{{\cite{%
nim:a277:176%
}}\xspace}
\def\citeSixmT{{\cite{%
thesis:gosau:2007%
}}\xspace}
\def\citeBAC{{\cite{%
nim:a300:480%
}}\xspace}
\def\citePCAL{{\cite{%
desy-92-066,*zfp:c63:391,*acpp:b32:2025%
}}\xspace}
\def\citeSPECTRO{{\cite{%
nim:a565:572%
}}\xspace}

\begin{document}
\graphicspath{{./}}
 
%
%
\prepnum{DESY--16--065}
\date{April 2016}
 
 
\title{
Search for a narrow baryonic state decaying to $\bvec{pK^0_S}$ and $\bvec{\overline{p}K^0_S}$ in deep inelastic scattering at HERA}
                    
\author{ZEUS Collaboration}
 
 
%
%
\abstract{
A search for a narrow baryonic state in the $pK^0_S$ and $\overline{p}K^0_S$ system has been performed in $ep$ collisions at HERA with the ZEUS detector using an integrated luminosity of $\unit{358}{pb^{-1}}$ taken in 2003--2007. The search was performed with deep inelastic scattering events at an $ep$ centre-of-mass energy of $\unit{318}{\GeV}$ for exchanged photon virtuality, $Q^2$, between 20 and
$\unit{100}{\GeV\squared}$. Contrary to evidence presented for such a state around \unit{1.52}{\GeV} in a previous ZEUS analysis using a sample of $121$~pb$^{-1}$ taken in 1996--2000, no resonance peak was found in the $p(\overline{p})K^0_S$ invariant-mass distribution in the range 1.45--\unit{1.7}{\GeV}. Upper limits on the production cross section are set.

}

\makezeustitle

\pagenumbering{roman}
\begin{center}
{                      \Large  The ZEUS Collaboration              }
\end{center}

{\small\raggedright


H.~Abramowicz$^{25, u}$, 
I.~Abt$^{20}$, 
L.~Adamczyk$^{8}$, 
M.~Adamus$^{31}$, 
S.~Antonelli$^{2}$, 
V.~Aushev$^{17}$, 
O.~Behnke$^{10}$, 
U.~Behrens$^{10}$, 
A.~Bertolin$^{22}$, 
S.~Bhadra$^{33}$
% ...
}
\newpage


{\setlength{\parskip}{0.4em}
\makebox[3ex]{$^{1}$}
\begin{minipage}[t]{14cm}
{\it INFN Bologna, Bologna, Italy}~$^{A}$

\end{minipage}

\makebox[3ex]{$^{2}$}
\begin{minipage}[t]{14cm}
{\it University and INFN Bologna, Bologna, Italy}~$^{A}$

\end{minipage}

\makebox[3ex]{$^{3}$}
\begin{minipage}[t]{14cm}
{\it Physikalisches Institut der Universit\"at Bonn,
Bonn, Germany}~$^{B}$

\end{minipage}

\makebox[3ex]{$^{4}$}
\begin{minipage}[t]{14cm}
{\it Panjab University, Department of Physics, Chandigarh, India}

\end{minipage}

\makebox[3ex]{$^{5}$}
\begin{minipage}[t]{14cm}
{\it Calabria University,
Physics Department and INFN, Cosenza, Italy}~$^{A}$

\end{minipage}

\makebox[3ex]{$^{6}$}
\begin{minipage}[t]{14cm}
{\it National Centre for Particle Physics, Universiti Malaya, 50603 Kuala Lumpur, Malaysia}~$^{C}$

\end{minipage}

\makebox[3ex]{$^{7}$}
\begin{minipage}[t]{14cm}
{\it The Henryk Niewodniczanski Institute of Nuclear Physics, Polish Academy of \\
Sciences, Krakow, Poland}~$^{D}$

\end{minipage}

\makebox[3ex]{$^{8}$}
\begin{minipage}[t]{14cm}
{\it AGH-University of Science and Technology, Faculty of Physics and Applied Computer
Science, Krakow, Poland}~$^{D}$

\end{minipage}

\makebox[3ex]{$^{9}$}
\begin{minipage}[t]{14cm}
{\it Department of Physics, Jagellonian University, Krakow, Poland}

\end{minipage}

\makebox[3ex]{$^{10}$}
\begin{minipage}[t]{14cm}
{\it Deutsches Elektronen-Synchrotron DESY, Hamburg, Germany}

\end{minipage}

\makebox[3ex]{$^{11}$}
\begin{minipage}[t]{14cm}
{\it Deutsches Elektronen-Synchrotron DESY, Zeuthen, Germany}

\end{minipage}

\makebox[3ex]{$^{12}$}
\begin{minipage}[t]{14cm}
{\it School of Physics and Astronomy, University of Glasgow,
Glasgow, United Kingdom}~$^{E}$

\end{minipage}

\makebox[3ex]{$^{13}$}
\begin{minipage}[t]{14cm}
{\it Hamburg University, Institute of Experimental Physics, Hamburg,
Germany}~$^{F}$

\end{minipage}

\makebox[3ex]{$^{14}$}
\begin{minipage}[t]{14cm}
{\it Institute of Particle and Nuclear Studies, KEK,
Tsukuba, Japan}~$^{G}$

\end{minipage}

\makebox[3ex]{$^{15}$}
\begin{minipage}[t]{14cm}
{\it Institute of Physics and Technology of Ministry of Education and
Science of Kazakhstan, Almaty, Kazakhstan}

\end{minipage}

\makebox[3ex]{$^{16}$}
\begin{minipage}[t]{14cm}
{\it Institute for Nuclear Research, National Academy of Sciences, Kyiv, Ukraine}

\end{minipage}

\makebox[3ex]{$^{17}$}
\begin{minipage}[t]{14cm}
{\it Department of Nuclear Physics, National Taras Shevchenko University of Kyiv, Kyiv, Ukraine}

\end{minipage}

\makebox[3ex]{$^{18}$}
\begin{minipage}[t]{14cm}
{\it Meiji Gakuin University, Faculty of General Education,
Yokohama, Japan}~$^{G}$

\end{minipage}

\makebox[3ex]{$^{19}$}
\begin{minipage}[t]{14cm}
{\it Lomonosov Moscow State University, Skobeltsyn Institute of Nuclear Physics,
Moscow, Russia}~$^{H}$

\end{minipage}

\makebox[3ex]{$^{20}$}
\begin{minipage}[t]{14cm}
{\it Max-Planck-Institut f\"ur Physik, M\"unchen, Germany}

\end{minipage}

\makebox[3ex]{$^{21}$}
\begin{minipage}[t]{14cm}
{\it Department of Physics, University of Oxford,
Oxford, United Kingdom}~$^{E}$

\end{minipage}

\makebox[3ex]{$^{22}$}
\begin{minipage}[t]{14cm}
{\it INFN Padova, Padova, Italy}~$^{A}$

\end{minipage}

\makebox[3ex]{$^{23}$}
\begin{minipage}[t]{14cm}
{\it Dipartimento di Fisica e Astronomia dell' Universit\`a and INFN,
Padova, Italy}~$^{A}$

\end{minipage}

\makebox[3ex]{$^{24}$}
\begin{minipage}[t]{14cm}
{\it Polytechnic University, Tokyo, Japan}~$^{G}$

\end{minipage}

\makebox[3ex]{$^{25}$}
\begin{minipage}[t]{14cm}
{\it Raymond and Beverly Sackler Faculty of Exact Sciences, School of Physics, \\
Tel Aviv University, Tel Aviv, Israel}~$^{I}$

\end{minipage}

\makebox[3ex]{$^{26}$}
\begin{minipage}[t]{14cm}
{\it Department of Physics, Tokyo Institute of Technology,
Tokyo, Japan}~$^{G}$

\end{minipage}

\makebox[3ex]{$^{27}$}
\begin{minipage}[t]{14cm}
{\it Universit\`a di Torino and INFN, Torino, Italy}~$^{A}$

\end{minipage}

\makebox[3ex]{$^{28}$}
\begin{minipage}[t]{14cm}
{\it Universit\`a del Piemonte Orientale, Novara, and INFN, Torino,
Italy}~$^{A}$

\end{minipage}

\makebox[3ex]{$^{29}$}
\begin{minipage}[t]{14cm}
{\it Physics and Astronomy Department, University College London,
London, United Kingdom}~$^{E}$

\end{minipage}

\makebox[3ex]{$^{30}$}
\begin{minipage}[t]{14cm}
{\it Faculty of Physics, University of Warsaw, Warsaw, Poland}

\end{minipage}

\makebox[3ex]{$^{31}$}
\begin{minipage}[t]{14cm}
{\it National Centre for Nuclear Research, Warsaw, Poland}

\end{minipage}

\makebox[3ex]{$^{32}$}
\begin{minipage}[t]{14cm}
{\it Department of Particle Physics and Astrophysics, Weizmann
Institute, Rehovot, Israel}

\end{minipage}

\makebox[3ex]{$^{33}$}
\begin{minipage}[t]{14cm}
{\it Department of Physics, York University, Ontario, Canada M3J 1P3}~$^{J}$

\end{minipage}

}

\vspace{3em}


{\setlength{\parskip}{0.4em}\raggedright
\makebox[3ex]{$^{ A}$}
\begin{minipage}[t]{14cm}
 supported by the Italian National Institute for Nuclear Physics (INFN) \
\end{minipage}

\makebox[3ex]{$^{ B}$}
\begin{minipage}[t]{14cm}
 supported by the German Federal Ministry for Education and Research (BMBF), under
 contract No. 05 H09PDF\
\end{minipage}

\makebox[3ex]{$^{ C}$}
\begin{minipage}[t]{14cm}
 supported by HIR grant UM.C/625/1/HIR/149 and UMRG grants RU006-2013, RP012A-13AFR and RP012B-13AFR from
 Universiti Malaya, and ERGS grant ER004-2012A from the Ministry of Education, Malaysia\
\end{minipage}

\makebox[3ex]{$^{ D}$}
\begin{minipage}[t]{14cm}
 supported by the National Science Centre under contract No. DEC-2012/06/M/ST2/00428\
\end{minipage}

\makebox[3ex]{$^{ E}$}
\begin{minipage}[t]{14cm}
 supported by the Science and Technology Facilities Council, UK\
\end{minipage}

\makebox[3ex]{$^{ F}$}
\begin{minipage}[t]{14cm}
 supported by the German Federal Ministry for Education and Research (BMBF), under
 contract No. 05h09GUF, and the SFB 676 of the Deutsche Forschungsgemeinschaft (DFG) \
\end{minipage}

\makebox[3ex]{$^{ G}$}
\begin{minipage}[t]{14cm}
 supported by the Japanese Ministry of Education, Culture, Sports, Science and Technology
 (MEXT) and its grants for Scientific Research\
\end{minipage}

\makebox[3ex]{$^{ H}$}
\begin{minipage}[t]{14cm}
 supported by RF Presidential grant N 3042.2014.2 for the Leading Scientific Schools\
\end{minipage}

\makebox[3ex]{$^{ I}$}
\begin{minipage}[t]{14cm}
 supported by the Israel Science Foundation\
\end{minipage}

\makebox[3ex]{$^{ J}$}
\begin{minipage}[t]{14cm}
 supported by the Natural Sciences and Engineering Research Council of Canada (NSERC) \
\end{minipage}

}

\pagebreak[4]
{\setlength{\parskip}{0.4em}


\makebox[3ex]{$^{ a}$}
\begin{minipage}[t]{14cm}
now at INFN Roma, Italy\
\end{minipage}

\makebox[3ex]{$^{ b}$}
\begin{minipage}[t]{14cm}
now at Sant Longowal Institute of Engineering and Technology, Longowal, Punjab, India\
\end{minipage}

\makebox[3ex]{$^{ c}$}
\begin{minipage}[t]{14cm}
now at Sri Guru Granth Sahib World University, Fatehgarh Sahib, India\
\end{minipage}

\makebox[3ex]{$^{ d}$}
\begin{minipage}[t]{14cm}
also at Agensi Nuklear Malaysia, 43000 Kajang, Bangi, Malaysia\
\end{minipage}

\makebox[3ex]{$^{ e}$}
\begin{minipage}[t]{14cm}
partially supported by the Polish National Science Centre projects DEC-2011/01/B/ST2/03643 and DEC-2011/03/B/ST2/00220\
\end{minipage}

\makebox[3ex]{$^{ f}$}
\begin{minipage}[t]{14cm}
now at Rockefeller University, New York, NY 10065, USA\
\end{minipage}

\makebox[3ex]{$^{ g}$}
\begin{minipage}[t]{14cm}
now at European X-ray Free-Electron Laser facility GmbH, Hamburg, Germany\
\end{minipage}

\makebox[3ex]{$^{ h}$}
\begin{minipage}[t]{14cm}
now at University of Liverpool, United Kingdom\
\end{minipage}

\makebox[3ex]{$^{ i}$}
\begin{minipage}[t]{14cm}
now at Tel Aviv University, Isreal\
\end{minipage}

\makebox[3ex]{$^{ j}$}
\begin{minipage}[t]{14cm}
now at Physikalisches Institut, Universit\"{a}t Heidelberg, Germany\
\end{minipage}

\makebox[3ex]{$^{ k}$}
\begin{minipage}[t]{14cm}
supported by the Alexander von Humboldt Foundation\
\end{minipage}

\makebox[3ex]{$^{ l}$}
\begin{minipage}[t]{14cm}
now at CERN, Geneva, Switzerland\
\end{minipage}

\makebox[3ex]{$^{ m}$}
\begin{minipage}[t]{14cm}
Alexander von Humboldt Professor; also at DESY and University of Oxford\
\end{minipage}

\makebox[3ex]{$^{ n}$}
\begin{minipage}[t]{14cm}
also at DESY\
\end{minipage}

\makebox[3ex]{$^{ o}$}
\begin{minipage}[t]{14cm}
also at University of Tokyo, Japan\
\end{minipage}

\makebox[3ex]{$^{ p}$}
\begin{minipage}[t]{14cm}
now at Kobe University, Japan\
\end{minipage}

\makebox[3ex]{$^{ q}$}
\begin{minipage}[t]{14cm}
member of National Technical University of Ukraine, Kyiv Polytechnic Institute, Kyiv, Ukraine\
\end{minipage}

\makebox[3ex]{$^{ r}$}
\begin{minipage}[t]{14cm}
now at DESY CMS group\
\end{minipage}

\makebox[3ex]{$^{ s}$}
\begin{minipage}[t]{14cm}
now at DESY ATLAS group\
\end{minipage}

\makebox[3ex]{$^{ t}$}
\begin{minipage}[t]{14cm}
now at LNF, Frascati, Italy\
\end{minipage}

\makebox[3ex]{$^{ u}$}
\begin{minipage}[t]{14cm}
also at Max Planck Institute for Physics, Munich, Germany, External Scientific Member\
\end{minipage}

\makebox[3ex]{$^{ v}$}
\begin{minipage}[t]{14cm}
also supported by DESY and the Alexander von Humboldt Foundation\
\end{minipage}

\makebox[3ex]{$^{ w}$}
\begin{minipage}[t]{14cm}
also at \L\'{o}d\'{z} University, Poland\
\end{minipage}

\makebox[3ex]{$^{ x}$}
\begin{minipage}[t]{14cm}
member of \L\'{o}d\'{z} University, Poland\
\end{minipage}

\makebox[3ex]{$^{ y}$}
\begin{minipage}[t]{14cm}
now at Polish Air Force Academy in Deblin\
\end{minipage}

}

\cleardoublepage
\cleardoublepage
 
\pagenumbering{arabic}
\section{Introduction}
\label{sec-int}
\indent
 The observation of a narrow baryon resonance with a mass of $\approx$\unit{1.53}{\GeV}, reported first by the LEPS  
 experiment in 2003~\cite{prl:91:012002,pr:c79:025210} in the missing-mass distribution for $\gamma A$ collisions, generated considerable
theoretical and experimental interest. Such a baryon would be manifestly exotic because of its decay into a $K^+$ and a neutron, which is impossible for a  three-quark state but could be explained as a bound state of five quarks i.e.\ a pentaquark state. A narrow baryonic resonance close to the observed mass had previously been predicted in the chiral soliton model \cite{zfp:a359:305} and named $\Theta^+$ with quark configuration $uudd\overline{s}$.  
Many experimental groups have looked for this state via various production processes in the decay modes $nK^+$ or $pK^0_S (\overline{p}K^0_S)$. Some experiments confirmed the signal while others refuted it. Several reviews \cite{Dzierba:2005,Danilov:2008,Hicks:2005a,Hicks:2005b,Hicks:2012} have been published on the subject.
 
The HERA accelerator collided electrons\footnote{In this paper, the word ``electron" refers to both electrons and positrons, unless otherwise stated.} at $E_e$ $=$ $\unit{27.5}{\GeV}$ with protons at $E_p$ $=$ 820 or $\unit{920}{\GeV}$. The ZEUS experiment reported evidence for a peak structure in the $pK^0_S$ mass distribution\footnote{Charge conjugated modes are implied throughout this paper, unless otherwise stated.} in deep inelastic scattering (DIS) data, consistent with a $\Theta^+$. The data were taken between 1996 and 2000 (HERA\,I) and correspond to an integrated luminosity of $121$~pb$^{-1}$~\cite{pl:b591:7}. The H1 collaboration presented mass distributions in a similar kinematic region~\cite{pl:b639:202}, but did not find any structure and presented an upper limit. However, this limit did not unambiguously  exclude the ZEUS signal.
 
 Recently, interest in pentaquark states has arisen again with the discovery of two pentaquark candidates by the LHCb experiment at 
$4.38$ and $4.45$~\GeV. They have a valence quark content of $uudc\overline{c}$ and were observed with high statistical significance~\cite{prl:115:72001}.
 
To clarify the production of strange pentaquarks in DIS, a search for the $\Theta^+$ resonance in the HERA\,II data (2003--2007) with an integrated luminosity of $358$~pb$^{-1}$ has been performed. The HERA\,II period not only provided larger statistics, but also the ZEUS tracking system was upgraded. In particular, a silicon-strip micro vertex detector (MVD)~\citeMVD located close to the beam line provided more information on the ionisation energy loss per unit length, $dE/dx$. This improves the selection of protons from a huge background of mainly pions. 
 
This paper presents the result of a search at HERA\,II for a narrow resonance in the $pK^0_S$ system in the central
 rapidity region of high-energy $ep$ collisions in a similar kinematic region to the previous ZEUS analysis. The sample includes both $e^+p$ and $e^-p$ collisions at a centre-of-mass energy of $\unit{318}{\GeV}$. The analysis was done with DIS events, requiring a visible scattered electron in the detector, at a photon virtuality, $Q^2$, in the range 20--\unit{100}{\GeV\squared}.
 
\section{Experimental set-up}
\label{sec-exp}
\Zdetdesc
 
Charged particles were tracked in the central tracking detector (CTD)~\citeCTD, the MVD~\citeMVD and the straw-tube tracking detector (STT)~\citeSTT. These components operated in a magnetic field of \unit{1.43}{\tesla} provided by a thin superconducting solenoid. The
CTD consisted of 72~cylindrical drift-chamber layers, organised in nine
superlayers covering the polar-angle\ZcoosysfnBEeta{} region
\mbox{$15^{\circ}<\theta<164^{\circ}$}.
The MVD silicon tracker consisted of a barrel (BMVD) and a forward
(FMVD) section. The BMVD contained three layers with two detectors in each layer and provided
polar-angle coverage for tracks from $30^{\circ}$ to
$150^{\circ}$. The four-layer FMVD extended the polar-angle coverage in
the forward region to $7^{\circ}$. The single-hit
resolution of the MVD was $\unit{24}{\mu \rm{m}}$. 
The transverse distance of closest
approach (DCA) of tracks to the nominal vertex in the $X$--$Y$ plane was measured to have
a resolution, averaged over the azimuthal angle, of                         
 ${(46 \oplus      122 / p_{T})}~{\mu \rm{m}}$, with $p_{T}$ in GeV.
For CTD--MVD tracks that pass through all nine CTD superlayers, the momentum resolution was
$\sigma(p_{T})/p_{T} = 0.0029 \, p_{T} \oplus  0.0081 \oplus 0.0012/p_{T}$, with $p_{T}$ in \GeV. Both the CTD and MVD were equipped with analog read-out systems which provided $dE/dx$ information for particle identification. The STT covered the polar-angle region $5^{\circ} < \theta < 25^{\circ}$.
 
\Zcaldesc
 
The luminosity was measured using the Bethe--Heitler reaction
$ep\,\rightarrow\, e\gamma p$ by a luminosity detector which consisted
of independent lead--scintillator calorimeter\citePCAL and magnetic
spectrometer\citeSPECTRO systems. The fractional systematic
uncertainty on the measured luminosity was 2\%~\cite{nim:a744:80}.

\section{Monte Carlo simulation}
\label{sec:monte-carlo}
\indent 
Samples of Monte Carlo (MC) events were generated to determine the detector acceptance in order to estimate the   
production cross section of a resonance state in the $pK^0_S$ system. The generated events were passed through the GEANT 3.21-based~\cite{tech:cern-dd-ee-84-1} ZEUS detector- and trigger-simulation programs~\cite{zeus:1993:bluebook}.  They were reconstructed and analysed by the same program chain as used for real data. 
 
Signal events were generated with the MC package RAPGAP v.3.1030~\cite{cpc:86:147}. Pentaquarks were simulated by replacing 
 $\Sigma^+(1189)$ in the particle table with a pentaquark with various masses (1.450, 1.500, 1.522, 1.540, 1.560, 1.600 and 
 \unit{1.650}{\GeV}), isotropically decaying into $pK^0$. Events that satisfy $Q^2 > \unit{1}{\GeV\squared}$ and $|y_{pK^0}| < 2.5 $, where $y_{pK^0}$ is the rapidity of the $pK^0$ system, were kept and processed in the detector simulation. Thirty million events were produced with $M =$ 1.522 and 
  $M =$ \unit{1.540}{\GeV}, which are the peak positions of the ZEUS HERA\,I analysis~\cite{pl:b591:7} and the PDG value of 2006~\cite{jphys:g33:1}, respectively. Fifteen million events were produced for each of the other mass points.

\section{Event selection}
\subsection{Event sample}
\indent
A three-level trigger~\cite{zeus:1993:bluebook,prep:92:150b,nim:A580:1257} was used to select DIS events, requiring scattered electron candidates. In the offline reconstruction, the scattered electron candidates were identified from the pattern of energy deposits in the CAL~\citeCAL. The Bjorken scaling variable, $x$, as well as $y$ and $Q^2$, were reconstructed using the double-angle method~\cite{proc:hera:1991:23,proc:hera:1991:43} which uses the angle of the scattered electron and the angle calculated from the remaining particles.
Here, $y = Q^2/(s x)$ denotes the fraction of the incoming electron energy transferred to the proton in the proton rest frame and $s$ is the square of the  centre-of-mass energy of the $ep$ system.
 
The following requirements, similar to those in the HERA\,I analysis, were imposed to select the events for the DIS sample:
\begin{itemize}
\item $20 < Q^2 < \unit{100}{\GeV\squared}$; 
\item $E_{e'} > \unit{10}{\GeV}$, where $E_{e'}$ is the corrected energy of the scattered electron measured in the CAL;
\item $38 < \delta < \unit{60}{\GeV}$, where $\delta = \Sigma E_i (1 - \cos \theta_i)$, $E_i$ is the energy of the $i$th calorimeter cell, $\theta_i$ is its polar angle and the sum runs over all cells; 
\item $y_{e} < 0.95$, and $y_{\rm JB} > 0.04$, where $y_{e}$ and $y_{\rm JB}$ are the $y$ values calculated by the electron and Jacquet--Blondel (JB) method \cite{proc:epfacility:1979:391}, respectively; 
\item $\mid Z_{\text{vertex}} \mid < \unit{30}{\cm}$, where $Z_{\text{vertex}}$ is the vertex position along the $Z$-axis determined from the tracks. 
\end{itemize}
The requirement $Q^2$ $>$ $\unit{20}{GeV^2}$ was motivated by the HERA\,I analysis; the requirement $Q^2$ $<$ $\unit{100}{GeV^2}$ allows a direct comparison to the H1 limit~\cite{pl:b639:202}.

In order to check the sensitivity of the HERA\,II data to resonance searches, the well-known $\Lambda_c(2286)$ baryon was searched for in the $pK_S^0$ mass spectrum in DIS and also in a photoproduction event sample, $Q^2 \approx \unit{0}{\GeV^2}$. The photoproduction events were collected from various trigger streams~\cite{thesis} by requiring offline that no identified electron with energy $E_{e'} > \unit{4}{\GeV}$ and $y_{e} < 0.85$ was found in the CAL and by imposing a cut $0.2 < \delta / E_e  < 0.85$, where $E_e$ is the electron beam energy. 
The same $Z_{\text{vertex}}$ cut was imposed as in the DIS sample.
 
\subsection{$\bvec{K^0_S}$ selection}
\indent
Neutral strange $K^0_S$ mesons were reconstructed from two charged tracks in the decay $K^0_S \rightarrow \pi^+ \pi^-$.  
The tracks were required to pass through at least three inner superlayers of the CTD, to have at least three BMVD hits out of the nominal six hits, and to have transverse momentum $p_T > \unit{0.15}{\GeV}$ and $|\eta| < 1.75$, restricting the study to a region where the track acceptance and momentum resolution were high. In view of the huge combinatorial background, only oppositely charged pairs whose three-dimensional distance of closest approach to each other was less than $\unit{1.5}{\cm}$ were considered for a vertex constraint fit. 
The invariant mass, $M(\pi^+\pi^-)$, was calculated assigning the $\pi$ mass to both tracks. The candidate pairs were required to satisfy the following conditions:
 
\begin{itemize}
\item $\chi^{2} < 5.0$, where $\chi^{2}$ refers to the re-fit of $K^0_S$ vertex position; 
\item $L_{XY}$ $>$ \unit{0.5}{\cm}, where $L_{XY}$ is the $K^0_S$ decay length in the $XY$ plane, to eliminate a background of misidentified decays close to the primary vertex;
\item $\alpha_{2D}$ $<$ 0.06 radian and $\alpha_{3D}$ $<$ 0.15 radian, where $\alpha_{2D}$ and $\alpha_{3D}$ are $XY$-projected and three-dimensional collinearity angles, respectively, defined as the angle between the direction from the primary vertex to the decay vertex and the momentum direction of the $\pi\pi$ system;
\item $p_T(K^0_S) > \unit{0.25}{\GeV}$, $\mid \eta(K^0_S) \mid < 1.6$.
\end{itemize}
 
In addition, the following requirements were imposed to eliminate contamination from  other sources:
\begin{itemize}
\item $M(e^+e^-) > \unit{0.07}{\GeV}$, where the electron mass was assigned to each track, to eliminate track pairs from photon conversions; 
\item $M(p\pi) > \unit{1.121}{\GeV}$, where the proton mass was assigned to the track with the higher momentum, to eliminate $\Lambda$ contamination of the $K^0_S$ signal.  
\end{itemize}
 
Figure~\ref{DISEV_K} shows the invariant-mass distribution for $K^0_S$ candidates. A fit with two Gaussian functions plus a constant was used.
The peak position was $M(K^0_S) = \unit{0.4972}{\GeV}$, which is consistent with the PDG value of \unit{0.4976}{\GeV}~\cite{chin:c38:090001} within the uncertainty on the momentum scale of the tracks (0.3\%).
The candidates with $0.482$ $<$ $M(\pi^+\pi^-)$ $<$ $\unit{0.512}{\GeV}$ were selected. A sample of 0.31 million events was selected with at least one $K^0_S$ candidate.
 
\subsection{Proton selection and particle identification}
\indent
The selection of proton or anti-proton tracks makes use of kinematic requirements and particle identification (PID). In the following, the term ``proton" denotes generically both the proton ($p$) and the anti-proton ($\overline{p}$). The kinematic selections on the proton track were as follows:
 
\begin{itemize}
	\item it passes through at least three inner superlayers of the CTD and has at least two MVD hits;
	\item its momentum, $p_{\rm{track}}$, satisfies 0.2 $<$ $p_{\rm{track}}$ $<$ \unit{1.5}{\GeV};
	\item it is associated with the primary vertex;
	\item it is not one of the tracks from the selected $K^0_S$ candidate.
\end{itemize}
 
The proton PID was performed with the combination of the CTD and MVD $dE/dx$ information. The $dE/dx$ in the CTD was estimated with the truncated-mean method used in previous ZEUS analyses~\cite{pl:b481:213,epj:c18:625}. 
The $dE/dx$ in the MVD was estimated by a likelihood method~\cite{thesis}.
The measured $dE/dx$ resolution was $\approx$10\% for each detector. 

The first step in selecting well measured protons required the measured $dE/dx$ values to be within bands centred at the expectation of the respective parameterised Bethe--Bloch function~\cite{chin:c38:090001}, and to be greater than 1.15 in units of minimum-ionising particles (mips). 
These cut positions are indicated in \figref{DISEV_P}, which shows CTD and MVD $dE/dx$ measurements as a function of $p_{\rm{track}}$.

The CTD and MVD $dE/dx$ measurements for the tracks selected as protons by the other detector are shown in \figsref{DISEV_P} (a) and (b), respectively. In addition to the clear proton bands, contaminations from kaons and pions are visible. In some cases, the CTD $dE/dx$ for tracks with large energy loss is not measured due to saturation of the signal; therefore there are fewer entries at high $dE/dx$ in the CTD plot (\figref{DISEV_P}(a)). 

In the second step, a likelihood-like estimator was used to select protons based on distances to the predicted Bethe--Bloch lines for proton, kaon and pion hypotheses.
In cases when the CTD $dE/dx$ was not determined because of a saturated signal, protons were selected using only the MVD $dE/dx$. Figures 2 (c) and (d) show the CTD and MVD $dE/dx$ distributions for tracks after the final selection. 
 
The proton identification efficiency of the $dE/dx$ selection was measured with a $\Lambda$ sample, selected using the $p\pi$ invariant mass without $dE/dx$ selection, from an extended DIS\footnote{In the extended DIS sample, no explicit $Q^2$ cut was imposed in order to keep as many $\Lambda$ candidates as possible.} sample and the photoproduction sample. The efficiency is about 80\% for protons with momentum $p_{\rm{track}} < \unit{0.8}{\GeV}$, almost linearly decreasing to 20\% at $p_{\rm{track}}=\unit{1.5}{\GeV}$, mainly due to the likelihood-like cut used to reduce the pion contamination. The identification efficiency for the protons from $\Lambda$ decays integrated over $p_{\rm{track}}$ from 0.1 to \unit{1.5}{\GeV} is 54\%. The pion-rejection factor was examined using pion tracks from $K^0_S$ decays. The factor is above 1000 for momenta below \unit{1.2}{\GeV} and decreases to 100 at \unit{1.5}{\GeV}.
 
For a direct comparison with the HERA\,I analysis, another event sample was prepared with protons selected using only the CTD $dE/dx$ using the first step of logic as described above. This results in a higher integrated proton identification efficiency of 82\% for protons in the $\Lambda$-decay sample, but the pion rejection factor above \unit{0.6}{\GeV}, where the increase in efficiency originates, is 10--100 times worse.

\section{Results}
\subsection{The $\bvec{pK^0_S}$ invariant-mass distribution} 
\indent
The $pK^0_S$ invariant mass was obtained by combining proton and $K^0_S$ candidates selected as described above and with their masses adjusted to the PDG value~\cite{chin:c38:090001}. The $pK^0_S$ candidates were selected in the kinematic region $0.5 < p_T(pK^0_S) < \unit{3.0}{\GeV}$ and $|\eta(pK^0_S)| < 1.5$.
 
The $pK^0_S$ invariant-mass distribution in the range from 1.4 to \unit{2.4}{\GeV} is shown in \figsref{DISEV_MASS} (a) and (b) for the DIS sample with $20 < Q^2 < \unit{100}{\GeV^2}$ and for the photoproduction sample. To suppress the combinatorial background for the $\Lambda_c(2286)$ production in the photoproduction sample, a requirement of $p_T(pK^0_S)$ $>$ $0.15 \, E_{T}^{\rm{}\theta>10^\circ}$ was imposed, where $E_{T}^{\rm{}\theta>10^\circ}$ is the sum of the transverse energy of the CAL cells outside a 10 degree cone from the proton-beam direction. This cut was motivated by the hard character of charm fragmentation.
 
A clear $\Lambda_c(2286)$ peak is observed in the photoproduction sample. It is also seen in the DIS sample with less significance. The width of the $\Lambda_c$ peak is \unit{10}{\MeV} and is consistent with the MC simulation. 
 
In \figref{DISEV_MASS}(c), the $pK^0_S$ invariant-mass distribution is shown in the mass range from 1.4 to \unit{1.9}{\GeV} for the same DIS sample with finer bins. The distribution contains 3107 $pK^0_S$ candidates and 2833 $\overline{p}K^0_S$ candidates. The pion contamination in the proton candidates was estimated to be less than 10\%.
The dashed line represents the $\Theta^+$ signal as would be observed if it had the same strength as reported in the ZEUS HERA\,I result.
The HERA\,I signal is not confirmed in this analysis.
 
For a more direct comparison of the present to the previous ZEUS result, an analysis with CTD-only $dE/dx$ selection and with similar cuts as in the HERA\,I analysis was performed. For this, no MVD information was used for the track selection. At least 40 CTD hits were required for the proton track. The result is shown in \figref{DISEV_MASS}(d).
The increase of the number of $pK^0_S$ candidates in \figref{DISEV_MASS}(d), of an order of magnitude with respect to \figref{DISEV_MASS}(c), is mainly due to the looser PID selection for the proton candidates. It is consistent with the number of candidates observed in the HERA\,I analysis.
For this looser selection, the pion contamination in the proton candidates was estimated to be more than 50\%. No peak is seen in \figref{DISEV_MASS}(d).
 
\subsection{Upper limits on the production cross section}
\indent
Since there is no significant structure in the invariant-mass distribution, upper limits on the production cross section of a narrow $pK^0$ resonance were derived.  

A fit was performed to the mass plot shown in \figref{DISEV_MASS}(c) for a mass range between 1.435 and \unit{1.9}{\GeV} with a Gaussian function for a postulated signal and an empirical function for background of the form
\begin{equation} 
\alpha (M-M_0)^{\beta} (1+\gamma (M-M_0)), \nonumber
\end{equation}
where $\alpha$, $\beta$ and $\gamma$ are parameters determined in the fit, $M$ is the $pK_S^0$ mass, and $M_0$ is the sum of the nominal proton and $K^0$ masses~\cite{chin:c38:090001}. 

Three options were used for the width of the Gaussian. One option was to fix it to $\unit{6.1}{\MeV}$, which is the measured value from the ZEUS HERA\,I analysis. In the other two options, the width was set to $1\times$ and $2\times$ the detector resolution.
The resolution of the $pK^0_S$ invariant mass was estimated using the MC events and was \unit{3.5}{\MeV} in the region near \unit{1.52}{\GeV} and \unit{11}{\MeV} near \unit{2.3}{\GeV}. For the mass range shown in \figref{DISEV_MASS}(c), the resolution $R$ was parameterised with the following formula;
\begin{equation}
R = 0.00959 \, M - 0.01111~{\rm (\GeV)}. \label{fwidth}
\end{equation}

The upper limit on the cross section at 95\% confidence level (CL) was determined at the value which increases the $\chi^2$ of the fit by 2.71~\cite{chin:c38:090001} with respect to the best fit\footnote{The best fit is obtained in the non-negative region of the signal amplitude. When the best-fit amplitude is zero, this gives a more conservative limit than at 95\% CL.}. At $M$ $=$ \unit{1.52}{\GeV}, where the peak was found in the HERA\,I analysis~\cite{pl:b591:7}, the obtained upper limit is 25.8 events for a width of \unit{6.1}{\MeV}.
For the HERA\,I analysis, ZEUS reported $221 \pm 48$ events above the background.
Correcting this number of events for the luminosity and for differences in the event selection and detector efficiencies, dominated by the proton identification, the predicted number of events for this analysis is 286. In \figref{DISEV_MASS}(c), a peak of this magnitude with resolution \unit{6.1}{\MeV} is shown as the dashed line above a solid curve which represents the background-only fit. Since no peak is observed at \unit{1.52}{\GeV}, the structure in the HERA\,I data is assumed to be a background fluctuation.

The cross sections were defined in the following kinematic range reflecting the region of large acceptance:
 
\begin{itemize}
\item $ 20$ $<$ $Q^2$ $<$ $\unit{100}{\GeV\squared}$;
\item $ | \eta(pK^0) |$ $<$ $1.5 $;
\item $ 0.5$ $<$ $p_T(pK^0)$ $<$ $\unit{3.0}{\GeV}$.
\end{itemize}
 
The final results are shown as upper limits to the production cross section for either $\Theta^+$ or $\overline{\Theta^+}$, multiplied by the branching ratio of $\Theta^+ \rightarrow p K^0$, i.e.
\begin{equation}
\sigma ({\Theta}) = ( \sigma(ep \rightarrow e \Theta^+ X) + \sigma(ep \rightarrow e \overline{\Theta^+} X)) \times BR(\Theta^+ \rightarrow p K^0). \nonumber
\end{equation}
The branching ratios of the $K^0$ to $K^0_S$ transition and of the $K^0_S$ to $\pi^+ \pi^-$ decay used in the cross-section calculation were 0.5 and 0.6895~\cite{chin:c38:090001} respectively.
 
The acceptance for the event selection was estimated using cross-section calculations from the MC samples except for the proton PID efficiency, which was determined from the $\Lambda$ sample. It was assumed that the $p_T$ and $\eta$ distributions of the resonance are similar to the $\Sigma^{\pm}(1189)$ as generated in RAPGAP v.3.1030~\cite{cpc:86:147} and that the resonance decays isotropically to $pK^0_S$. Since the detection efficiency depends strongly on the $(p_T,\eta)$ values of the $pK^0_S$ system, some variations on the $p_T$ distribution were tested as a study of the systematic uncertainty.
 
Systematic uncertainties on the cross section were evaluated for the following 4 components:
\begin {itemize} 
\item uncertainty in the event selection: the acceptance corrections were recalculated by shifting selection cuts~\cite{thesis} and re-evaluating the upper limit on the cross section. The variance was about 10\%;
\item the proton PID efficiency was modified by $\pm 1 \sigma$ of the measurement uncertainty. The effect was about 3\% with little mass dependence; 
\item uncertainty in the mass-dependent selection efficiency: the acceptance for a $pK^0_S$ resonance was determined using the seven MC samples for different masses as defined in Section 3. The mass dependence of the efficiency was fitted with a linear or a quadratic function to obtain the value for any given mass.  
The difference between the two fit functions gave a negligible contribution to the systematic uncertainty; 
\item model uncertainty on the $p_T$ distribution of a $pK^0_S$ resonance: in this analysis, the MC samples were generated using RAPGAP by replacing  $\Sigma^{\pm}(1189)$ with resonant states at various masses (see Section 3). In the model, the $p_T$ distribution was less steep with increasing mass. As a test, the distribution was re-scaled in order to keep the same $p_T$ spectra for all masses. At high masses, this gave about 20\% difference. 
\end{itemize}
In addition, there was a 2\% uncertainty on the luminosity measurement~\cite{nim:a744:80}. All resulting variations on the upper limit of the cross sections were added in quadrature and the upper limit was increased accordingly.
 
The upper limits\footnote{Since in the present analysis the origin of $K^0_S$ from $K^0$ or $\overline{K^{0}}$ cannot be distinguished, all limits are equally valid for a hypothetical narrow $p\overline{K^{0}}$ resonance.} obtained on $\sigma ({\Theta})$ at 95\% CL are shown in \figref{SUM_LIMIT}(a) for a width of the $\Theta^+$ of \unit{6.1}{\MeV}. As a reference, the limit considering only the statistical uncertainty is also shown. The limit in the region of the $\Theta^+$ mass is below \unit{10}{pb}. 
 
In \figref{SUM_LIMIT}(b), the cross-section limits for a $\Theta^+$ with an intrinsic width much smaller than the detector resolution (see \eref{fwidth}) is shown. Also shown are the limits for a $\Theta^+$ with a width reconstructed as twice the detector resolution, which approximately corresponds to the width used for the published H1 limit. The ZEUS limit is more stringent than that obtained by H1.

\section{Summary}
\indent
A resonance in the $pK^0_S(\overline{p}K^0_S)$ system consistent with a $\Theta^+$-like state has been searched for in the HERA\,II data  collected with the ZEUS detector, exploiting the improved proton identification capability made possible by the use of the micro vertex detector. A peak at \unit{1.52}{\GeV} for which evidence had been observed in a previous ZEUS analysis, based on HERA\,I data, was not confirmed.   
Upper limits on the production cross section of such a resonance have been set as a function of the $pK^0$ mass in the kinematic region:   
 $0.5<p_T(pK^0)<\unit{3.0}{\GeV}$, $|\eta(pK^0)| < 1.5$ and  $20 < Q^2 < \unit{100}{\GeV\squared}$.

\section*{Acknowledgements}
\label{sec-ack}

\Zacknowledge
 
\vfill\eject
 
\clearpage
{\raggedright
 
{
\ifzeusbst
  \ifzmcite
     \bibliographystyle{l4z_default3}
  \else
     \bibliographystyle{l4z_default3_nomcite}
  \fi
\fi
\ifzdrftbst
  \ifzmcite
    \bibliographystyle{l4z_draft3}
  \else
    \bibliographystyle{l4z_draft3_nomcite}
  \fi
\fi
\ifzbstepj
  \ifzmcite
    \bibliographystyle{l4z_epj3}
  \else
    \bibliographystyle{l4z_epj3_nomcite}
  \fi
\fi
\ifzbstjhep
  \ifzmcite
    \bibliographystyle{l4z_jhep3}
  \else
    \bibliographystyle{l4z_jhep3_nomcite}
  \fi
\fi
\ifzbstnp
  \ifzmcite
    \bibliographystyle{l4z_np3}
  \else
    \bibliographystyle{l4z_np3_nomcite}
  \fi
\fi
\ifzbstpl
  \ifzmcite
    \bibliographystyle{l4z_pl3}
  \else
    \bibliographystyle{l4z_pl3_nomcite}
  \fi
\fi
{\raggedright
\bibliography{l4z_zeus,%
              l4z_h1,%
              l4z_articles,%
              l4z_books,%
              l4z_conferences,%
              l4z_misc,%
              l4z_preprints}}
}
\vfill\eject

\begin{figure}
\begin{center}
\includegraphics[height=10cm,width=10cm]{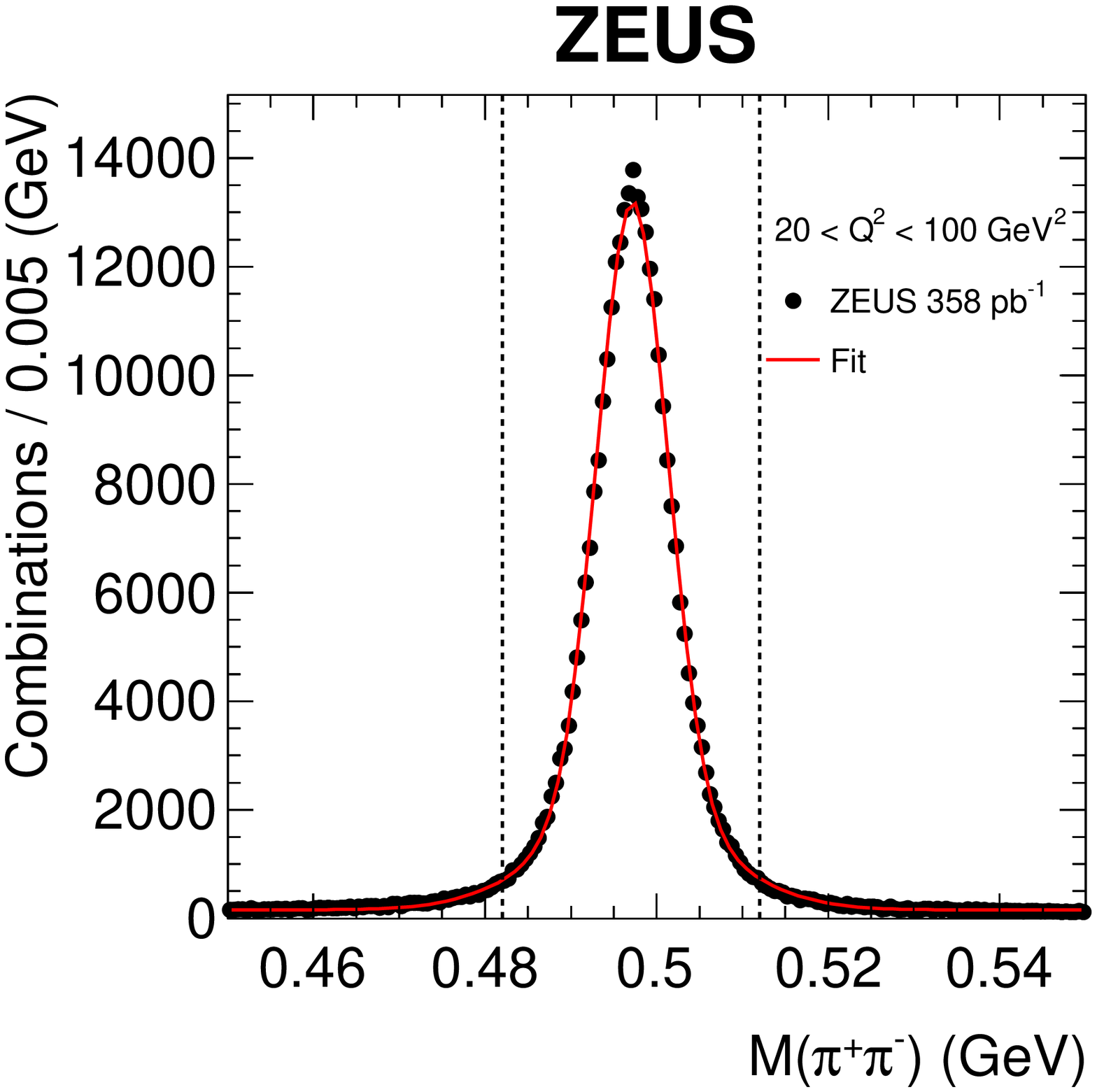}
\hangcaption{The $\pi^+\pi^-$ invariant-mass distribution for $20$ $<$ $Q^2$ $<$ $\unit{100}{\GeV\squared}$. The dashed lines show the mass range used for the $K^0_S$ selection. For illustration, the result of a fit with two Gaussian functions and constant background is shown. } \label{DISEV_K} 
\end{center}
\end{figure}

\begin{figure}
\begin{center}
\includegraphics[width=15cm]{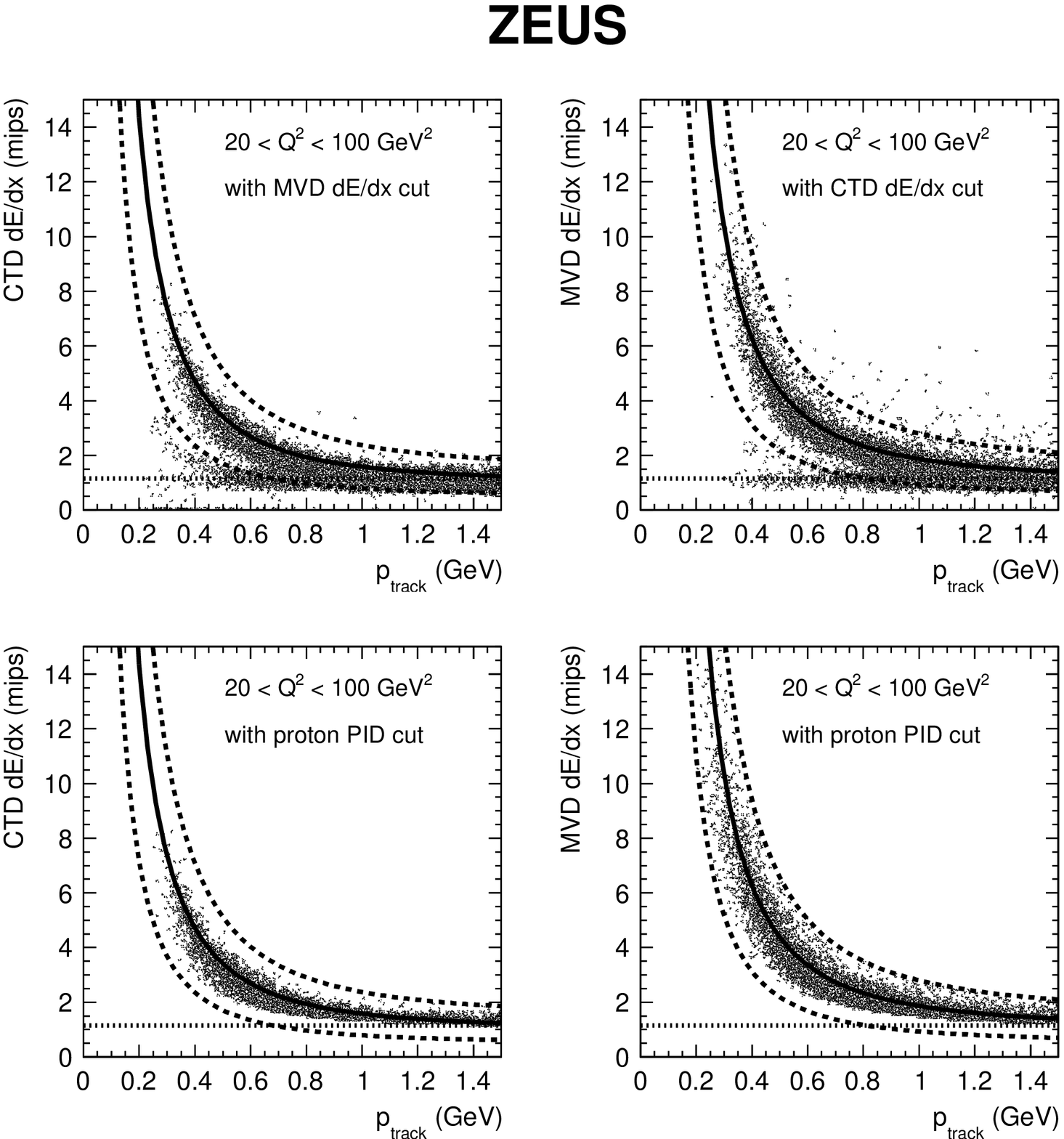}
\hspace*{10mm}
\\[-144.5mm]
\hspace{4.8cm}
(a)\hspace{7.0cm}(b)\\[6.8cm]
\hspace{4.8cm}
(c)\hspace{7.0cm}(d)\\[6.cm]
\hangcaption{The $dE/dx$ distributions as a function of $p_{\rm{track}}$ for (a) the CTD and (b) the MVD for the tracks identified as protons by the $dE/dx$ of the other detector; the distributions for (c) the CTD and (d) the MVD for the tracks finally selected as protons including tracks for which $dE/dx$ information was only available from the MVD. 
The solid lines show the Bethe--Bloch values for the proton. The dashed lines indicate the limits used for the proton selection. The dotted line is drawn at $1.15$ $mips$, the value used for the proton selection.} \label{DISEV_P} 
\end{center}
\end{figure}
 
\begin{figure}
\begin{center}
\includegraphics[width=15cm]{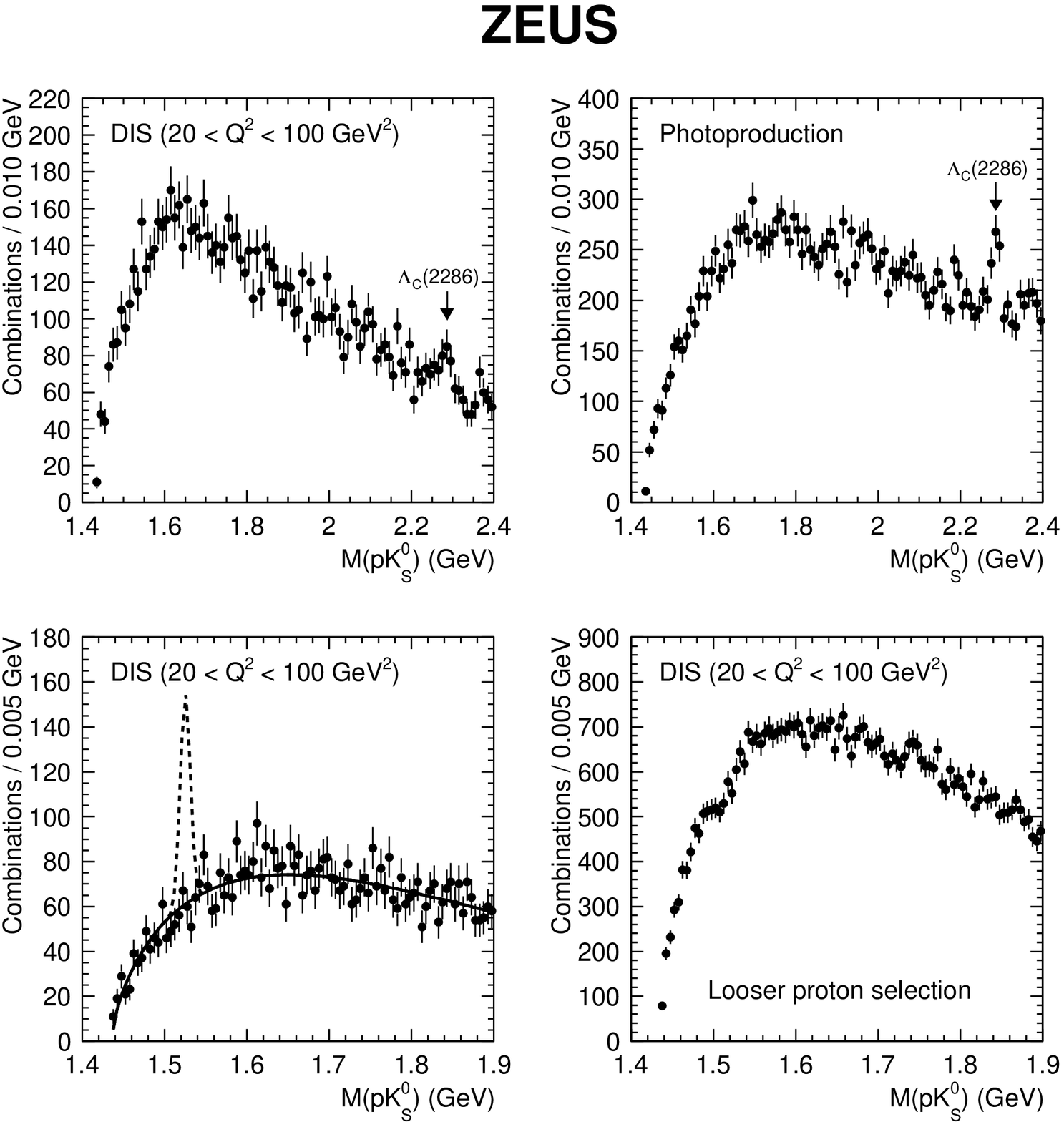}
\hspace*{10mm}
\\[-145.2mm]
\hspace{4.5cm}
(a)\hspace{7cm}(b)\\[6.8cm]
\hspace{4.5cm}
(c)\hspace{7cm}(d)\\[6.0cm]
\hangcaption{The $pK^0_S$ invariant-mass distribution for (a) the DIS sample with $20$ $<$ $Q^2$ $<$ $\unit{100}{\GeV^2}$ and (b) the  photoproduction sample. (c) The $pK^0_S$ distribution for the DIS sample with smaller bins.
The solid line is the result of a fit using the background function. The dashed line represents the signal corresponding to the ZEUS HERA\,I result. (d) The $pK^0_S$ distribution as in (c) with proton PID according to the HERA\,I analysis.} \label{DISEV_MASS} 
\end{center}
\end{figure}

\begin{figure}
\begin{center}
\includegraphics[width=15cm]{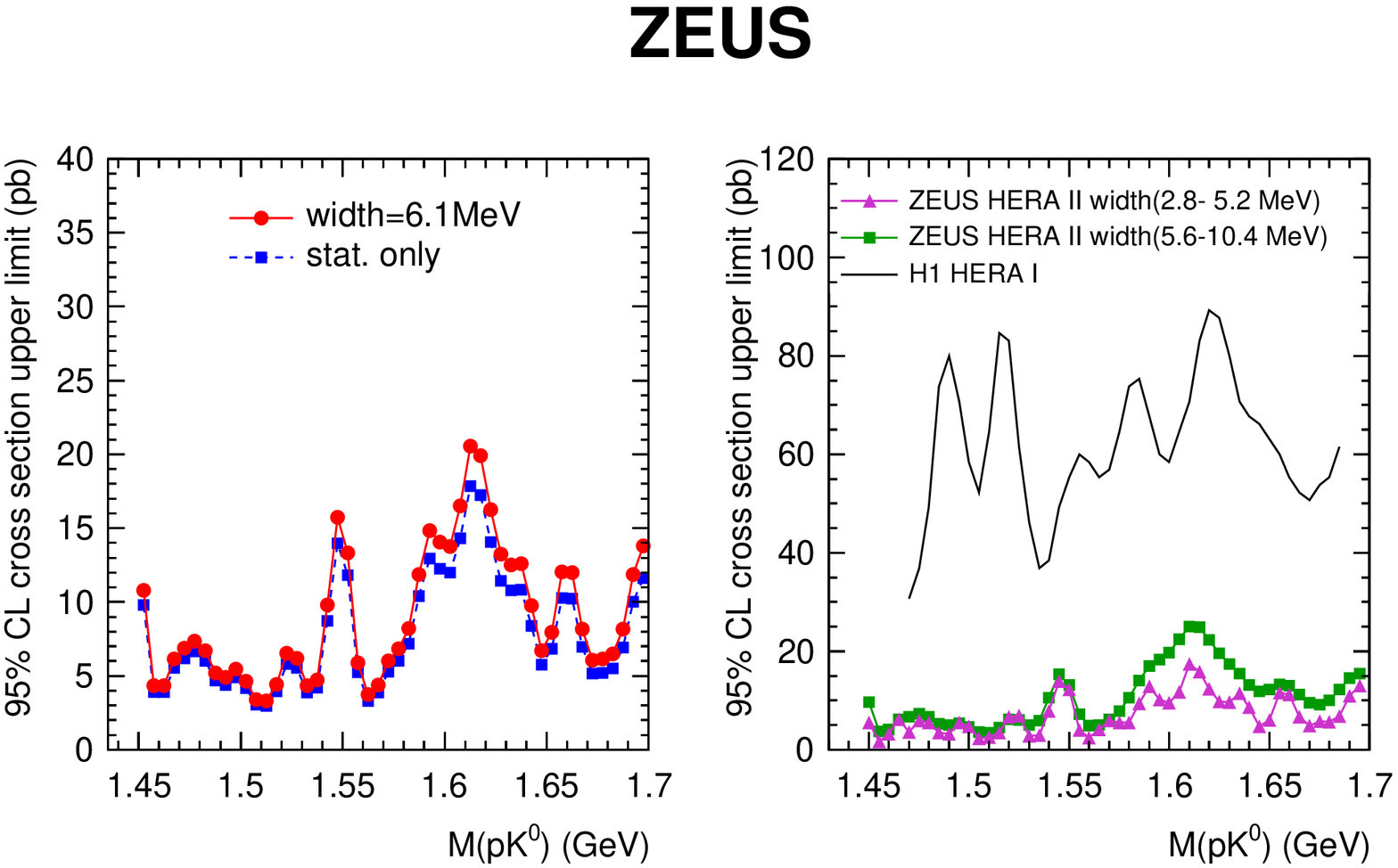}
\hspace*{10mm}
\\[-67mm]
\hspace{4.5cm}
(a)\hspace{7cm}(b)\\[6cm]
\hangcaption{The 95\% CL upper limits on $\sigma ({\Theta})$ for different hypotheses on the width of the observed peak;
(a) \unit{6.1}{\MeV} and (b) the mass resolution and twice the mass resolution.
In (a), the limit set by the statistical uncertainty only is also shown. In (b), the limit from the H1 result is also shown.} \label{SUM_LIMIT}
\end{center}
\end{figure}

}
 
\end{document}